# Interaction of degenerate higher order modes in periodic SRF accelerating structures

V. Yakovlev*, S. Belomestnykh, T. Khabiboulline, and A. Lunin
*Fermi National Accelerator Laboratory (FNAL), Kirk and Pine St., Batavia, IL 60510*

**ABSTRACT**. Some higher order mode (HOM) passbands in multicell SRF accelerating structures may become “flat,” meaning that they have an extremely narrow bandwidth due to weak coupling between cells. In the presence of cell misalignments, such passbands can split into individual cell resonances that are weakly coupled or completely uncoupled from the beam pipes. As a result, these modes can exhibit high quality factors and large *R/Q* values, potentially leading to beam instabilities, emittance growth, and excessive cryogenic losses. This effect was observed in a high-beta 5-cell 650 MHz (HB650) prototype cavity developed for the PIP-II project. An analytical theory explaining the origin of this phenomenon, along with methods to mitigate it, is presented in this paper. The HB650 cavity was successfully redesigned to eliminate the flat HOM passband and prevent the associated detrimental effects.

## I. INTRODUCTION.

The excitation of higher order modes (HOMs) in superconducting RF cavities is a well-known phenomenon that is routinely considered during cavity design. HOMs with high loaded quality factors and large *R/Q* values can be excited by the accelerated beam to significant amplitudes, potentially resulting in beam emittance dilution, excessive cryogenic load, and even beam instabilities. Particular attention must be paid to the so-called trapped modes; see, for example, [1]. These modes are characterized by electromagnetic fields concentrated in the central cells of a multi-cell cavity, with weak field amplitudes at the ends, resulting in high loaded quality factors and large *R/Q* values.

The excitation of HOMs has been extensively studied and is generally well understood. Dedicated conferences and numerous publications are devoted to this subject, and, at first glance, it may appear difficult to identify any new effects. However, during the development of the high-beta, 5-cell 650 MHz SRF cavity prototype [2] for the PIP-II project [3], an unexpected phenomenon was encountered. Instead of the expected fifth monopole HOM passband, a set of individual cell resonances with closely spaced frequencies around 1,988 MHz was observed [4].

The cavity layout and the results of axial field distribution measurements using the bead-pool technique are shown in Fig. 1, which demonstrates the presence of trapped monopole modes. Individual resonances localized in the central cells of the cavity are weakly coupled to the input coupler and, therefore, exhibit very high loaded quality factors. Since the electromagnetic fields of these resonances are predominantly confined within individual cells, the fields on the axis are also highly localized, resulting in a high transit time factor and, consequently, large *R/Q* values. Combined with a high loaded quality factor, this leads to a large shunt impedance.

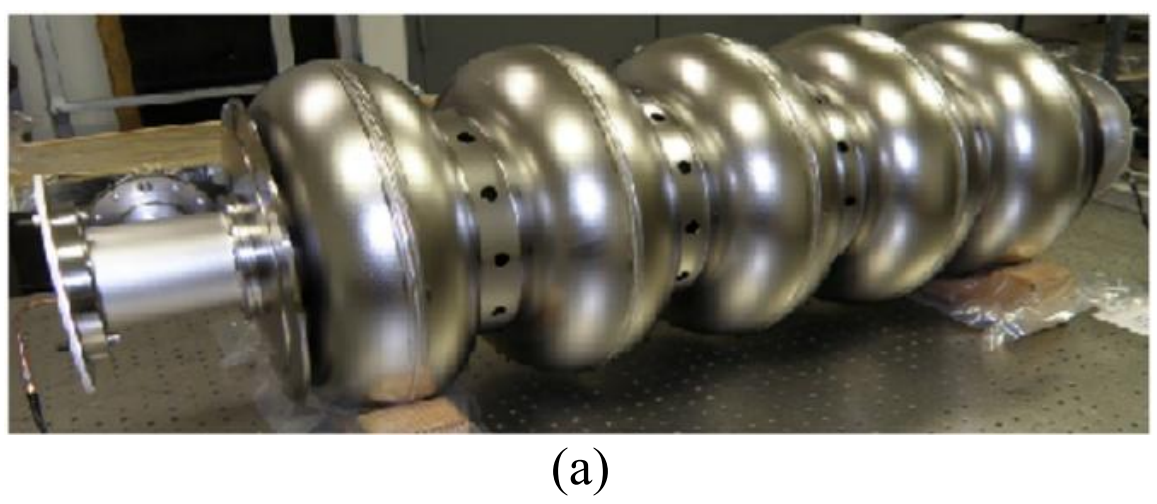

(a)

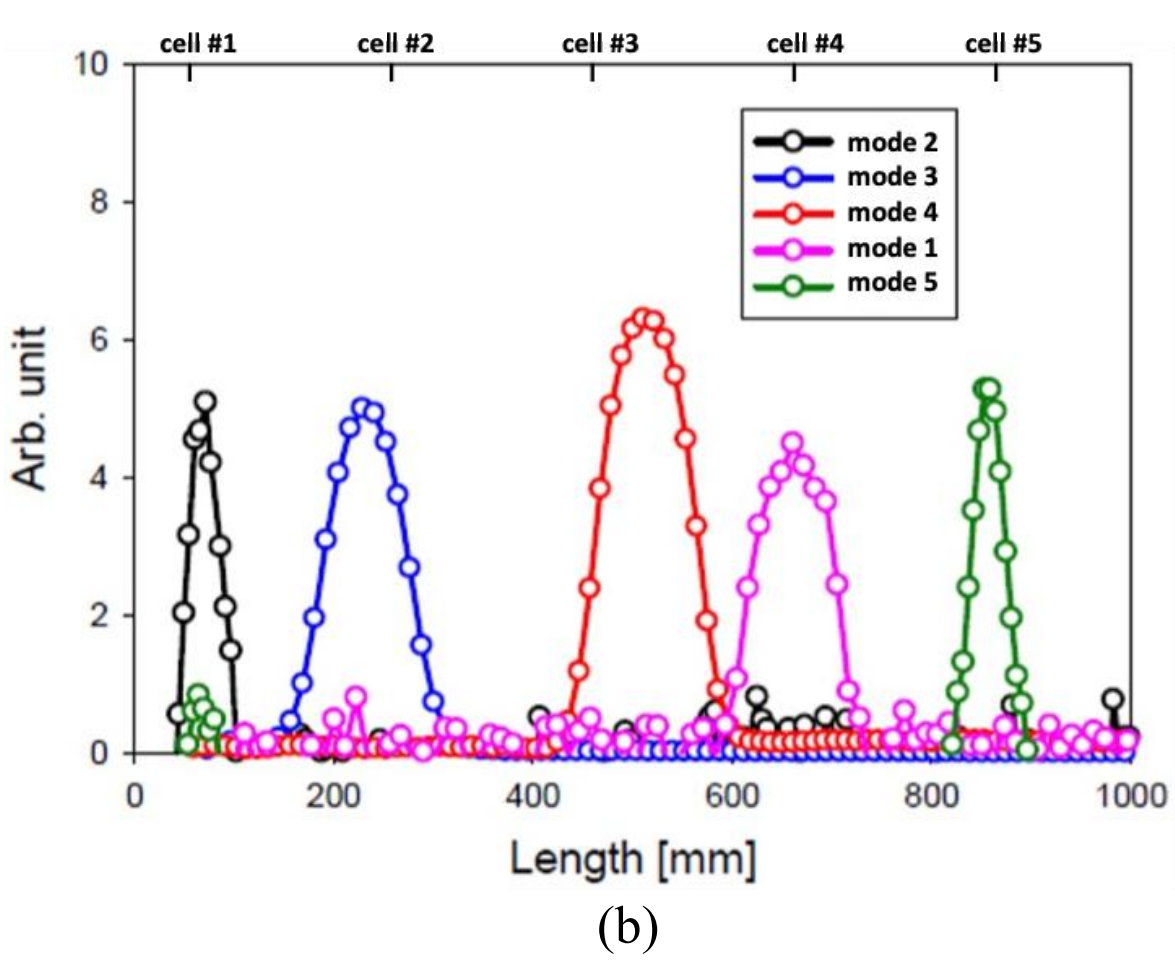

(b)

**Figure 1:** (a) Photograph of the high-beta, 5-cell 650 MHz SRF cavity prototype developed for the PIP-II project, with the input coupler port located on the left-hand side; and (b) bead-pull measurements of the on-axis field profiles for five monopole modes, showing excellent agreement with the prediction that, in a real cavity with cell misalignments, the passband splits into individual cell resonances.

*Contact author: yakovlev@fnal.gov

Results of simulations for an ideal cavity [5] show an extremely narrow passband with a bandwidth of approximately 40 kHz. The weak coupling between the cavity cells is caused by the small electric and magnetic fields in the vicinity of the cavity apertures. The corresponding electric field patterns in the cavity are shown in Fig. 2 (left). Due to weak coupling between the cells, even slight geometry r.m.s. perturbations of $\pm$ 0.2 mm can cause the passband to split into individual cell resonances [5] , as shown in Fig. 2 (right).

Attempts to increase the inter-cell coupling coefficient $K$ by enlarging the cavity aperture from the initial value of 50 mm to 60 mm did not significantly improve the situation. The coupling coefficient remains relatively insensitive to the aperture size until a certain critical value is reached. However, by combining the aperture increase with an increase in the cell length, i.e., increasing the cavity geometric beta $\beta_G$ from 0.9 to 0.92, it was possible to increase the inter-cell coupling coefficient to an acceptable level and eliminate the passband splitting [5]. As it will be shown in subsequent sections, the cell modes of the 5th passband are hybrid modes resulting from the coupling of degenerate modes, such as $TM_{030}$ and $TM_{012}$.

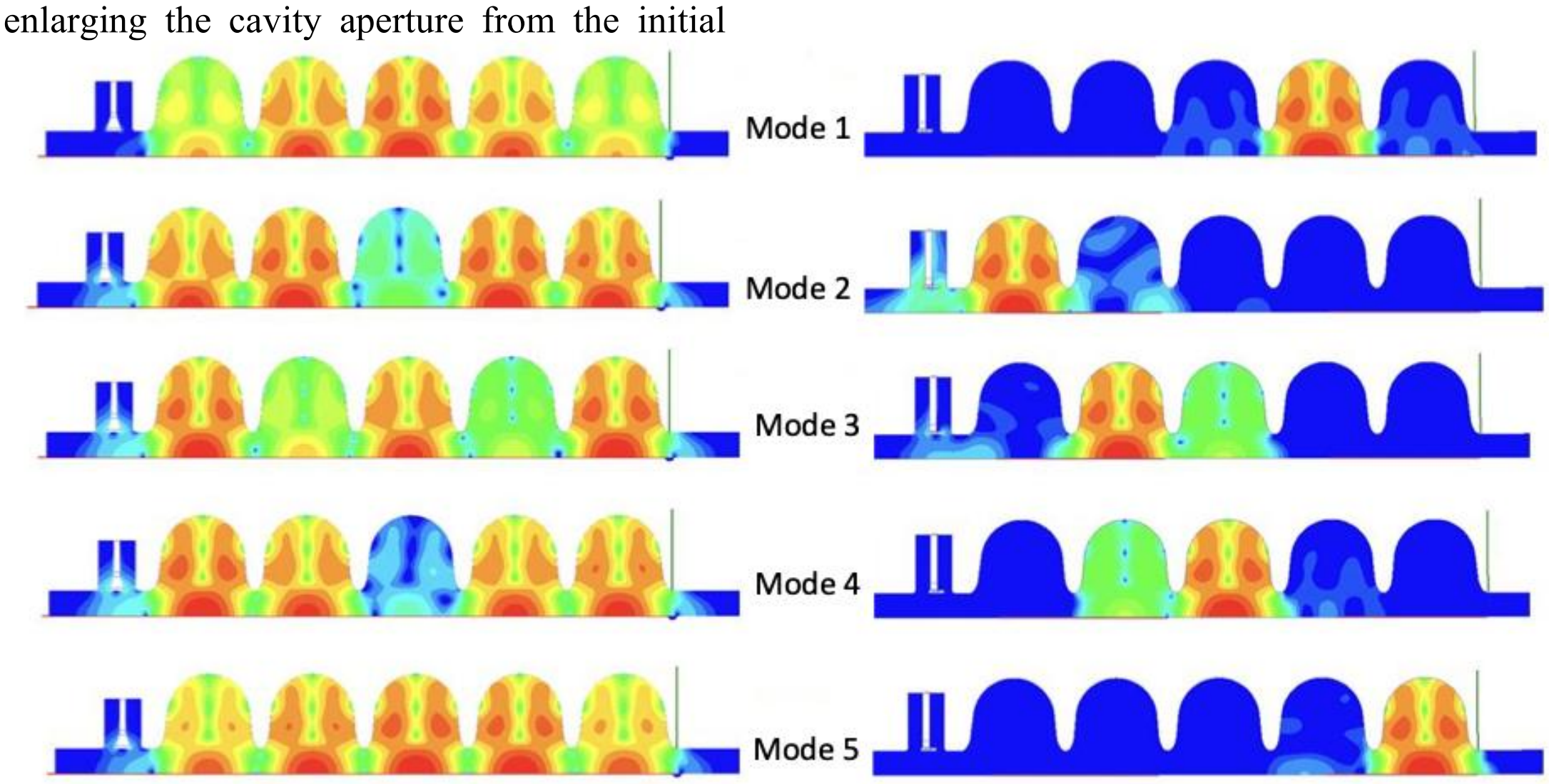

**Figure 2:** Electric field maps for hybrid modes of the 5th monopole passband in the initial design of a five-cell 650 MHz SRF cavity with the $\beta_G = 0.9$ for the PIP-II project: for the cases of an ideal cavity (left) and a cavity with r.m.s. misalignments of $\pm$ 0.2 mm (right).

## II. PERTURBATION THEORY FOR DEGENERATED MODES IN RF CAVITIES.

Perturbation theory, typically used to analyze multi-cell SRF cavities, works well in cases where the resonant frequencies of the passband modes do not coincide [6]. However, in certain cavities, like deflecting cavities [7] or travelling-wave cavities with feedback waveguide [8], degenerate modes exist for example, two dipole modes with different polarizations or partially orthogonal standing wave modes forming the travelling wave. In this case, linear perturbation theory does not apply, and coupled-mode theory is used. Below there is a brief outline of this approach to multi-cell SRF cavities based on the work [9], which details the effects of the mode interaction in RF cavities.

Consider *a concave* deformation of a cavity with two orthogonal modes, $\alpha$ and $\beta$, having close frequencies $\omega_\alpha$ and $\omega_\beta$ . The fields of these modes are $\vec{H}_\alpha$, $\vec{E}_\alpha$ and $\vec{H}_\beta$, $\vec{E}_\beta$ respectively. These fields satisfy ideal boundary conditions on the unperturbed cavity surface $S$. Figure 3 illustrates a cavity with a small concave deformation. The unperturbed cavity has volume $V$, and the volume of the perturbation is $\Delta V$. The field of a deformed cavity may be expressed, to a first approximation, via the fields of an ideal cavity as

$$\vec{\boldsymbol{H}} \approx h_\alpha \vec{H}_\alpha + h_\beta \vec{H}_\beta \tag{1}$$

$$\vec{\boldsymbol{E}} \approx e_\alpha \vec{E}_\alpha + e_\beta \vec{E}_\beta \tag{2}$$

with ideal boundary conditions on the deformed cavity surface $S_1$

$$\vec{n} \times \vec{\boldsymbol{E}} = 0, \tag{3}$$

$$\vec{n} \cdot \vec{\boldsymbol{H}} = 0. \tag{4}$$

To eliminate the field in the perturbation volume $\Delta V$, we introduce the *electric* currents $\vec{K}$ on the surface $S_1$, assuming that the tangential component of the electric field $\vec{\boldsymbol{E}}$ on the $S_1$ is zero, in accordance with (3). In this case, the magnetic field amplitudes $h_\alpha$ and $h_\beta$ of the modes are:

$$h_\alpha = \frac{\omega_\alpha \int_{S_1} \vec{K}\cdot\vec{E}_\alpha^* dS}{i(\omega^2-\omega_\alpha^2)\mu_0 \int_V \vec{H}_\alpha\cdot\vec{H}_\alpha^* dV}, \qquad (5)$$

$$h_\beta = \frac{\omega_\beta \int_{S_1} \vec{K}\cdot\vec{E}_\beta^* dS}{i(\omega^2-\omega_\beta^2)\mu_0 \int_V \vec{H}_\beta\cdot\vec{H}_\beta^* dV}. \qquad (6)$$

* Contact author: yakovlev@fnal.gov

Here $\omega$ is the resonance frequency of the deformed cavity, $\omega \approx \omega_\alpha \approx \omega_\beta$.

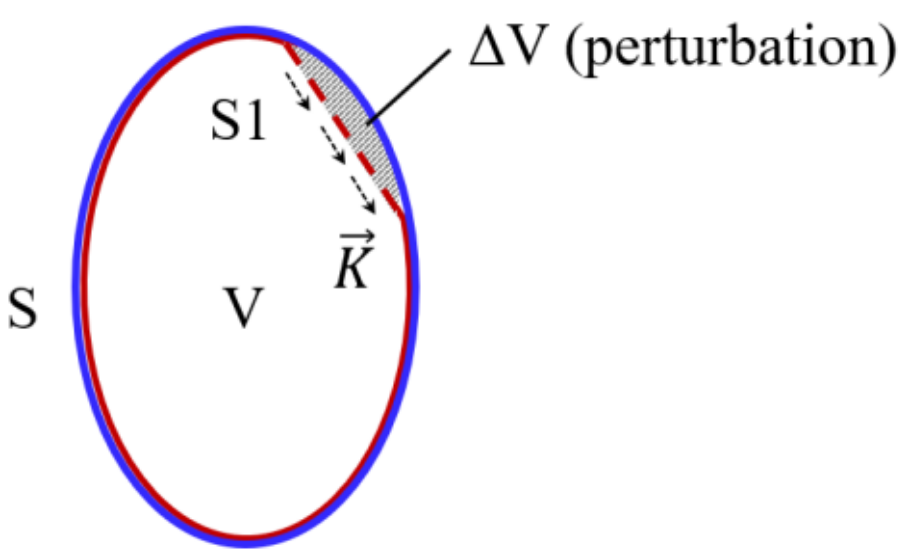

**Figure 3:** Concept of a cavity of volume $V$ with a small perturbation $\Delta V$; the unperturbed and perturbed geometries are shown by blue and red lines, respectively, along with the surface currents $\vec{K}$ on the interface surface $S_1$.

On the other hand, according to (1), on $S_1$ we have

$$\vec{K} = -\vec{n}\times\boldsymbol{H} \approx -h_\alpha(\vec{n}\times\vec{H}_\alpha) - h_\beta(\vec{n}\times\vec{H}_\beta). \qquad (7)$$

Substituting (7) into (5) and (6), using Maxwell's equations and relevant vector theorems, we get dispersion equations, which determine resonance frequencies of the two-mode system as well as the normal modes [9]:

$$\omega_\alpha^2 h_\alpha \left[1 + \frac{\int_{\Delta V}\left(\mu_0|\vec{H}_\alpha|^2 - \varepsilon_0|\vec{E}_\alpha|^2\right)dV}{\mu_0\int_V |\vec{H}_\alpha|^2 dV}\right] - \omega^2 h_\alpha +$$

$$+\omega_\alpha^2 h_\beta \frac{\int_{\Delta V}\left(\mu_0 \vec{H}_\alpha^*\cdot\vec{H}_\beta - \frac{\omega_\beta}{\omega_\alpha}\varepsilon_0\vec{E}_\alpha^*\cdot\vec{E}_\beta\right)dV}{\mu_0\int_V |\vec{H}_\alpha|^2 dV} = 0, \qquad (8)$$

$$\omega_\beta^2 h_\beta \left[1 + \frac{\int_{\Delta V}\left(\mu_0|\vec{H}_\beta|^2 - \varepsilon_0|\vec{E}_\beta|^2\right)dV}{\mu_0\int_V |\vec{H}_\beta|^2 dV}\right] - \omega^2 h_\beta +$$

$$+\omega_\beta^2 h_\alpha \frac{\int_{\Delta V}\left(\mu_0 \vec{H}_\beta^*\cdot\vec{H}_\alpha - \frac{\omega_\alpha}{\omega_\beta}\varepsilon_0\vec{E}_\beta^*\cdot\vec{E}_\alpha\right)dV}{\mu_0\int_V |\vec{H}_\beta|^2 dV} = 0. \qquad (9)$$

Equations (8) and (9) may be re-written the following way:

$$h_\alpha[\omega_\alpha^2(1+z_{\alpha\alpha}) - \omega^2] + h_\beta\omega_\alpha^2 z_{\alpha\beta} = 0, \quad (10)$$

$$h_\alpha\omega_\beta^2 z_{\beta\alpha} + h_\beta[\omega_\beta^2(1+z_{\beta\beta}) - \omega^2] = 0,$$

where

$$z_{\alpha\alpha} = \frac{\int_{\Delta V}\left(\mu_0|\vec{H}_\alpha|^2 - \varepsilon_0|\vec{E}_\alpha|^2\right)dV}{\mu_0\int_V |\vec{H}_\alpha|^2 dV},$$

$$z_{\beta\beta} = \frac{\int_{\Delta V}\left(\mu_0|\vec{H}_\beta|^2 - \varepsilon_0|\vec{E}_\beta|^2\right)dV}{\mu_0\int_V |\vec{H}_\beta|^2 dV}, \qquad (11)$$

$$z_{\alpha\beta} = \frac{\int_{\Delta V}\left(\mu_0 \vec{H}_\alpha^*\cdot\vec{H}_\beta - \frac{\omega_\beta}{\omega_\alpha}\varepsilon_0\vec{E}_\alpha^*\cdot\vec{E}_\beta\right)dV}{\mu_0\int_V |\vec{H}_\alpha|^2 dV},$$

$$z_{\beta\alpha} = \frac{\int_{\Delta V}\left(\mu_0 \vec{H}_\beta^*\cdot\vec{H}_\alpha - \frac{\omega_\alpha}{\omega_\beta}\varepsilon_0\vec{E}_\beta^*\cdot\vec{E}_\alpha\right)dV}{\mu_0\int_V |\vec{H}_\beta|^2 dV}.$$

The system of linear equations (10) can be represented in the matrix form,

$$\begin{pmatrix} \omega_\alpha^2(1+z_{\alpha\alpha}) - \omega^2 & \omega_\alpha^2 z_{\alpha\beta} \\ \omega_\beta^2 z_{\beta\alpha} & \omega_\beta^2(1+z_{\beta\beta}) - \omega^2 \end{pmatrix}\begin{pmatrix} h_\alpha \\ h_\beta \end{pmatrix} = 0 \quad (10')$$

The resonance frequencies correspond to the zeros of the matrix (10') determinant:

$$\begin{vmatrix} \omega_\alpha^2(1+z_{\alpha\alpha}) - \omega^2 & \omega_\alpha^2 z_{\alpha\beta} \\ \omega_\beta^2 z_{\beta\alpha} & \omega_\beta^2(1+z_{\beta\beta}) - \omega^2 \end{vmatrix} = 0. \quad (10'')$$

It is a secular equation for two coupled oscillators, e.g., see [10]. From this equation it follows that the resonance frequencies are given by well-known formula [9]:

$$\omega_\pm^2 = \frac{\omega_\alpha^2(1+z_{\alpha\alpha}) + \omega_\beta^2(1+z_{\beta\beta})}{2} \pm$$

$$\pm\sqrt{\left[\frac{\omega_\alpha^2(1+z_{\alpha\alpha}) - \omega_\beta^2(1+z_{\beta\beta})}{2}\right]^2 - \omega_\alpha^2\omega_\beta^2 z_{\alpha\beta}z_{\beta\alpha}}. \quad (12)$$

The amplitudes of the normal modes are the eigen vectors of the problem (10):

$$h_\alpha[\omega_\alpha^2(1+z_{\alpha\alpha}) - \omega_\pm^2] + h_\beta\omega_\alpha^2 z_{\alpha\beta} = 0 \qquad (13)$$

## III. MODE INTERACTION IN A PILLBOX CAVITY

To understand the phenomenon of mode interaction in a multi-cell periodic structure, we first consider an ideal multi-cell cavity composed of pillbox cells operating at the $TM_{010}$ mode and separated by thin diaphragms. The cell geometry is characterized by a radius $R$ and a length $d = \frac{\beta_G\lambda_{010}}{2}$, where $\beta_G$ is the so-called geometric beta of the periodic structure designed to accelerate particles with a velocity $\beta = v/c = \beta_G$. Pillbox cells are coupled via beam apertures of radius $a$ in the diaphragms. Let us first consider a separate cell. The higher order modes $TM_{012}$ and $TM_{020}$ in a pillbox cell have the following resonant frequencies:

* Contact author: yakovlev@fnal.gov

$$\omega_{012} = c\sqrt{\left(\frac{\nu_{01}}{R}\right)^2 + \left(\frac{\pi}{d}\right)^2}, \quad \omega_{020} = c\,\frac{\nu_{02}}{R} \qquad (14)$$

This pair of modes is degenerate at $\beta_G = 0.968$ , i.e., $\omega_{012} = \omega_{020} = \omega_0$. Let us model the beam apertures as axisymmetric coaxial protrusions on the end walls, each with a volume of $\Delta V/2$, and consider the mode mixing caused by these apertures. Note that in this case we have *convex* deformations, and therefore, the quantity ΔV should be taken *with the negative sign.* To prove this, one may use the same technique as in [9] but introduce *magnetic* currents $\vec{N}$ on the inner surface of the perturbation volume $\Delta V$. Since the aperture radius $a$ is small compared to the wavelength $\lambda_{012} = \lambda_{020}$, by setting $\vec{H}_\alpha = \vec{H}_{012}$ and $\vec{H}_\beta = \vec{H}_{020}$, one can use equation (1) for the total field.

For the partial modes under consideration

$$E_{\alpha z} = E_\alpha \cos\left(\frac{2\pi z}{d}\right) J_0\left(\frac{\nu_{01} r}{R}\right),$$

$$E_{\alpha r} = E_\alpha \frac{2\pi R}{\nu_{01} d} \sin\left(\frac{2\pi z}{d}\right) J_1\left(\frac{\nu_{01} r}{R}\right), \qquad (15)$$

$$H_{\alpha\varphi} = i\frac{E_\alpha}{Z_0}\frac{\nu_{02}}{\nu_{01}} \cos\left(\frac{2\pi z}{d}\right) J_1\left(\frac{\nu_{01} r}{R}\right),$$

and

$$E_{\beta z} = E_\beta J_0\left(\frac{\nu_{02} r}{R}\right),$$

$$E_{\beta r} = 0, \qquad (16)$$

$$H_{\beta\varphi} = i\frac{E_\beta}{Z_0} J_1\left(\frac{\nu_{02} r}{R}\right).$$

From (13) one has

$$h_\alpha\left[\omega_0^2(1 + \mathrm{Z}_{\alpha\alpha}) - \omega_\pm^2\right] + h_\beta \omega_0^2 Z_{\alpha\beta} = 0$$

and therefore,

$$h_\beta = -h_\alpha \frac{\omega_0^2(1+\mathrm{Z}_{\alpha\alpha}) - \omega_\pm^2}{\omega_0^2 Z_{\alpha\beta}}. \qquad (17)$$

Here

$$Z_{pq} \approx \frac{\int_{\Delta V}\left(\mu_0 \vec{H}_p^* \cdot \vec{H}_q - \varepsilon_0 \vec{E}_p^* \cdot \vec{E}_q\right) dV}{\mu_0 \int_V |\vec{H}_p|^2 dV}, \qquad (18)$$

$p$ and $q$ take the values $\alpha$ and $\beta$. For a small protrusion near the axis, a direct calculation using (11), (15) and (16) and neglecting magnetic field yields

$$Z_{\alpha\alpha} \approx 2\frac{\Delta V}{V}\left(\frac{\nu_{01}}{\nu_{02}}\right)^2 \frac{1}{J_1^2(\nu_{01})},$$

$$Z_{\beta\beta} \approx \frac{\Delta V}{V}\frac{1}{J_1^2(\nu_{02})}, \qquad (19)$$

$$Z_{\alpha\beta} \approx 2\frac{\Delta V}{V}\left(\frac{\nu_{01}}{\nu_{02}}\right)^2 \frac{1}{J_1^2(\nu_{01})}\frac{E_\beta}{E_\alpha} = Z_{\alpha\alpha}\frac{E_\beta}{E_\alpha},$$

$$Z_{\beta\alpha} \approx \frac{\Delta V}{V}\frac{1}{J_1^2(\nu_{02})}\frac{E_\alpha}{E_\beta} = Z_{\beta\beta}\frac{E_\alpha}{E_\beta}.$$

Note that we have taken into account that the perturbation is *convex*, which gives the positive sign of the quantity $Z_{pq}$ (see above).

From (12) and (19) one has

$$\omega_\pm^2 = \omega_0^2\left(1 + \frac{Z_{\alpha\alpha}+Z_{\beta\beta}}{2} \pm \frac{Z_{\alpha\alpha}+Z_{\beta\beta}}{2}\right). \qquad (20)$$

As follows from (15) and (16), the magnetic fields of the eigenmodes at radii $r < a$ are equal to

$$H_{\alpha\varphi} \approx i\frac{E_\alpha}{2Z_0}\frac{\nu_{02} r}{R}\cos\left(\frac{2\pi z}{d}\right), \qquad (21)$$

$$H_{\beta\varphi} \approx i\frac{E_\beta}{2Z_0}\frac{\nu_{02} r}{R}.$$

In this case, the total azimuthal magnetic field has the following form, taking into account Eq. (1):

$$H_\varphi(r,z) \approx i\frac{h_\alpha E_\alpha}{2Z_0}\frac{\nu_{02} r}{R}\left\{\cos\left(\frac{2\pi z}{d}\right) + \frac{1}{2}[(\eta-1) \pm (\eta+1)]\right\}, \quad (22)$$

here

$$\eta = \frac{1}{2}\left(\frac{\nu_{02} J_1(\nu_{01})}{\nu_{01} J_1(\nu_{02})}\right)^2 = 6.11.$$

As follows directly from the Maxwell's equation $\mathrm{curl}\,\vec{H} = i\omega_0\varepsilon_0\vec{E}$, the total longitudinal electric field on axis is

$$E_z(r,z) \approx h_\alpha E_\alpha\left\{\cos\left(\frac{2\pi z}{d}\right) + \frac{1}{2}[(\eta-1) \pm (\eta+1)]\right\}.$$

Let us examine these two modes in more detail.

a) For the "+" mode:

$$\omega_+^2 = \omega_0^2(1 + Z_{\alpha\alpha} + Z_{\beta\beta}) = \omega_0^2\left[1 + \frac{\Delta V}{V}\left(\left(\frac{\nu_{01}}{\nu_{02}}\right)^2 \frac{2}{J_1^2(\nu_{01})} + \frac{1}{J_1^2(\nu_{02})}\right)\right]$$

and the fields near the axis are:

$$E_z(r,z) \approx h_\alpha E_\alpha\left[\cos\left(\frac{2\pi z}{d}\right) + \eta\right],$$

$$H_\varphi(r,z) \approx i\frac{h_\alpha E_\alpha}{2Z_0}\frac{\nu_{02} r}{R}\left[\cos\left(\frac{2\pi z}{d}\right) + \eta\right].$$

b) For the "−" mode:

$$\omega_-^2 = \omega_0^2$$

and the fields near the axis are:

$$E_z(r,z) \approx h_\alpha E_\alpha\left[\cos\left(\frac{2\pi z}{d}\right) - 1\right], \qquad (24)$$

$$H_\varphi(r,z) \approx i\frac{h_\alpha E_\alpha}{2Z_0}\frac{\nu_{02} r}{R}\left[\cos\left(\frac{2\pi z}{d}\right) - 1\right].$$

On the cavity walls, i.e., when $z = 0$ or $z = d$, both electric and magnetic fields are zero. Thus, the distribution of the fields in the cavity is independent of the boundary conditions on the protrusion. This means that the cavity cannot be excited through apertures on the axis, and this is also a manifestation of the well-known resonance disappearance effect discovered for degenerate dipole modes [9, 11]. Figure 4 illustrates the effect of mode mixing for the degenerate modes discussed above, while Fig. 5 shows the normalized longitudinal component of the electric field on the axis for the “−” mode. The blue curve in Fig. 5 is calculated numerically for a pillbox cavity with protrusions (see Fig. 3), and the red one is calculated analytically using Eq. (24).

* Contact author: yakovlev@fnal.gov

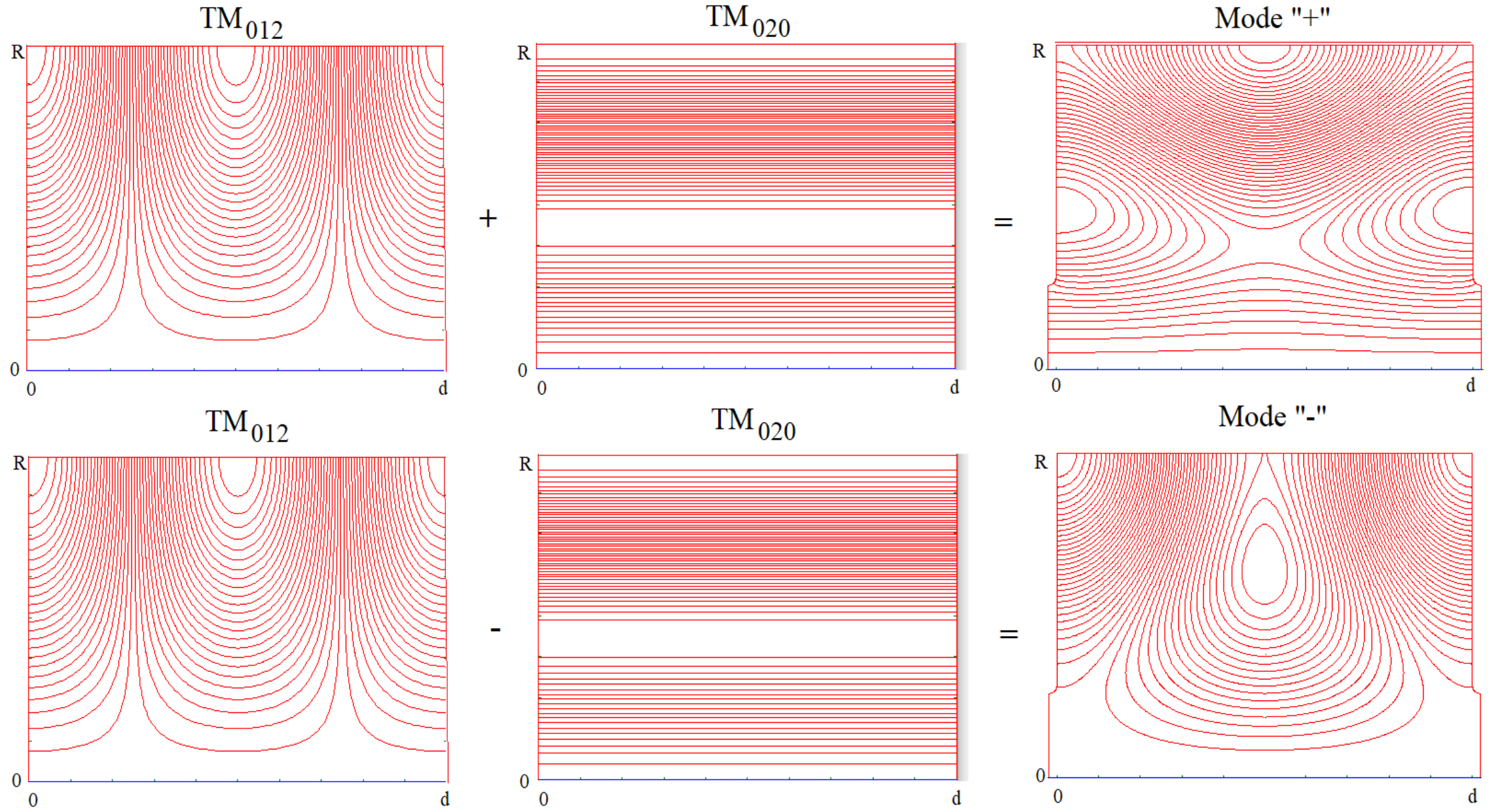

**Figure 4:** Mode interaction in a pillbox cavity. Left and center plots: electric field maps of unperturbed partial modes $TM_{012}$ and $TM_{020}$. Right plots: combined “+” mode (top plot) and “−” mode (bottom plot) in the presence of protrusions on the cavity end walls.

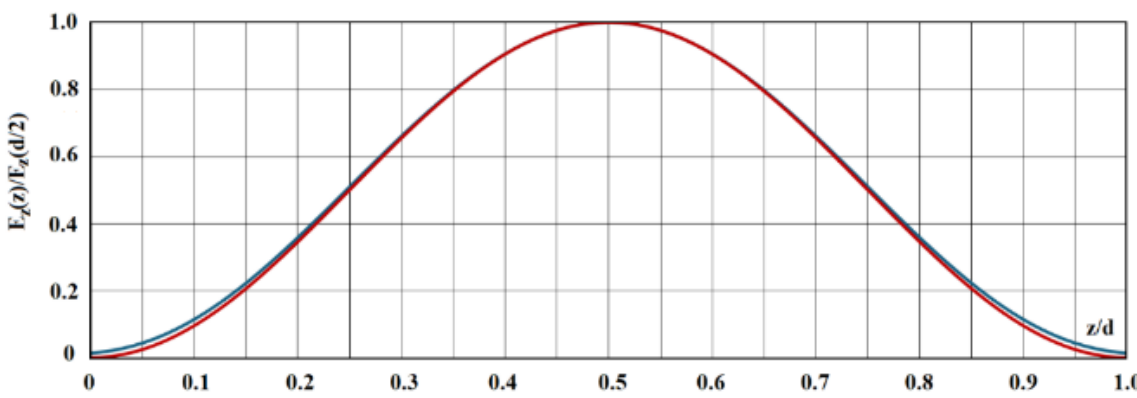

**Figure 5:** Axial distribution of the longitudinal electric field in a pillbox cavity. Blue curve: numerical simulation results for a pillbox cavity with protrusions. Red curve: analytical calculations according to Eq. (24).

Mode mixing, which leads to the vanishing electromagnetic field near the coupling apertures, can also be observed for other pairs of modes, see Fig. 6, which shows the calculated HOM frequencies in a pillbox cavity. The red dots correspond to the degeneracy between the $TM_{0n0}$ and $TM_{0m2}$ modes, which results in zero electromagnetic fields on the axis near the cavity end walls.

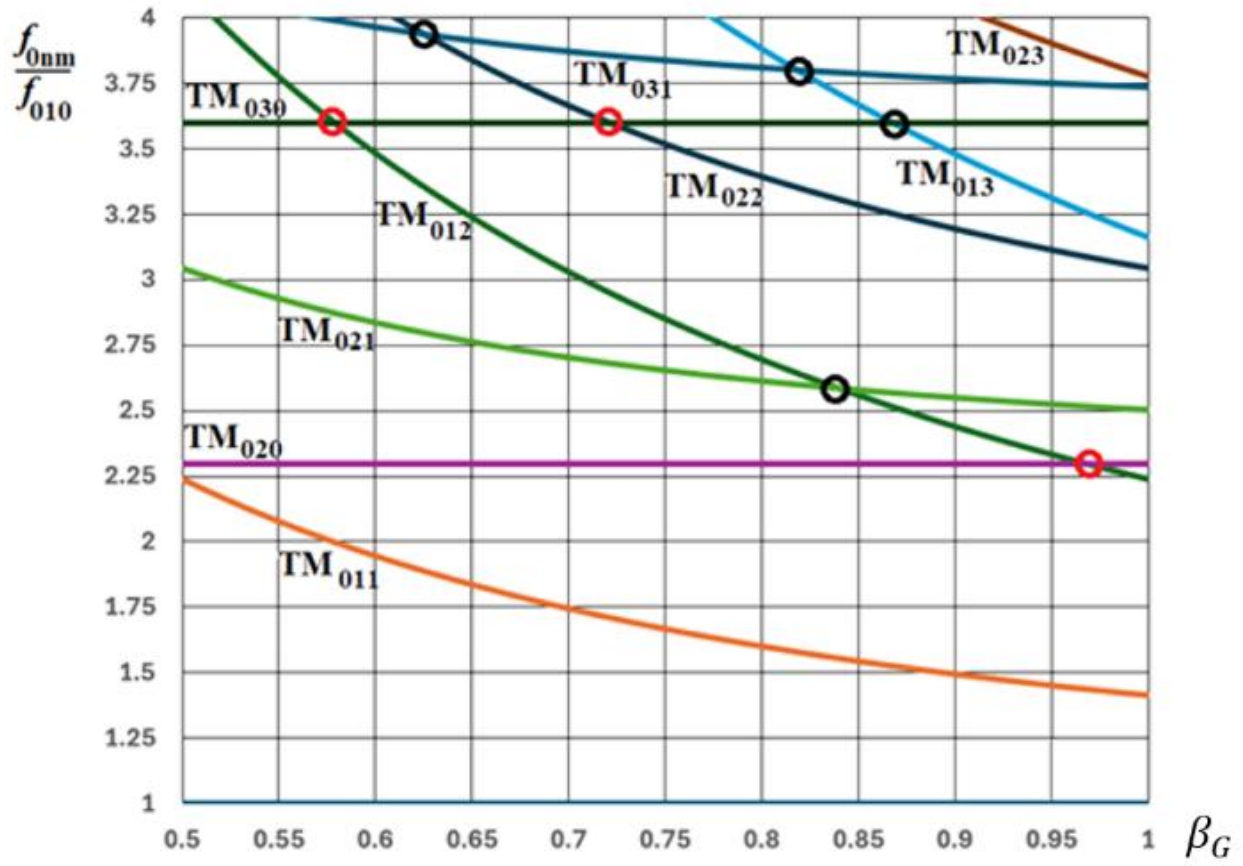

**Figure 6:** Normalized frequencies of the HOMs in a pillbox cavity as a function of parameter $\beta_G = d/(\lambda_{010}/2)$, where $d$ is the cavity length and $\lambda_{010}$ is the wavelength of the operating $TM_{010}$ mode.

For example, for the $TM_{030}$ and $TM_{022}$ modes, degeneracy occurs at $\beta_G = d/(\lambda_{010}/2) = 0.72168$. Consequently, if the $TM_{0n0}$ and $TM_{0m2}$ modes are degenerate, it is easy to see from Eq (14-24) that, in the first approximation, the electromagnetic field near the axis next to the cavity end walls is zero.

* Contact author: yakovlev@fnal.gov

Consider a multi-cell cavity comprised of pillbox cells coupled via apertures. For degenerate modes, both magnetic and electric fields on the axis near the end walls are small. In this case the inter-cell coupling coefficient (coupled-circuit model) is $K \ll 1$.[?] Fig. 7 illustrates the behavior of the coupling $K$ for two normal modes, "+" and "-", of such a periodic structure as a function of the normalized coupling aperture radius $a/R$. It can be seen that the coupling for the mode "$-$" is very weak up to $a/R \approx 0.45$, whereas the coupling for the mode "+" increases with the aperture as $\sim a^3$ [12] up to $a/R \approx 0.3$. Since for the hybrid mode "$-$", $K$ is very small, the frequencies of the multi-cell cavity in this passband are practically independent of the phase shift per cell. This implies that the passband is "flat". However, with a flat passband the field distribution for the normal standing-wave (SW) HOM mode along the multi-cell cavity becomes highly sensitive to perturbations in cell frequencies.

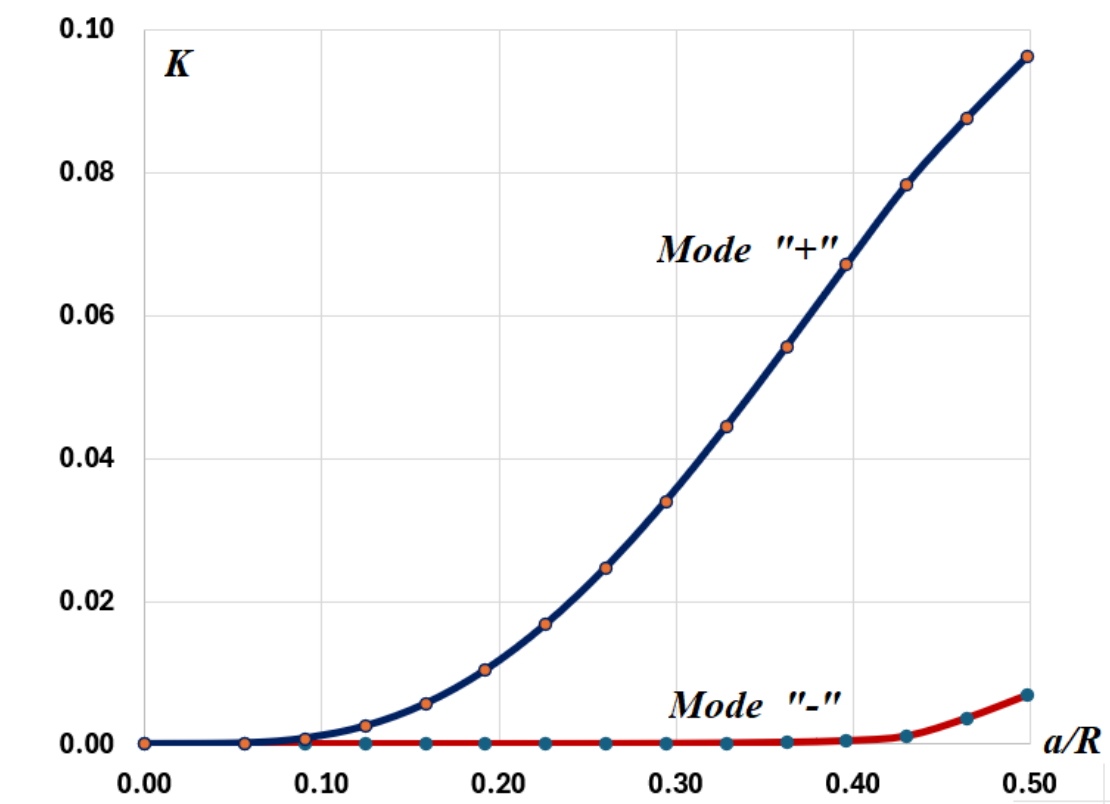

**Figure 7:** Coupling coefficient $K$ for the normal modes corresponding to the coupled $TM_{020}$ and $TM_{012}$ modes in a multi-cell pillbox cavity as a function of the normalized aperture radius.

According to [6], the relative field perturbation $\frac{\delta X}{X}$ of the SW mode in a multi-cell cavity is proportional to the relative perturbation of the cell frequency $\frac{\delta\omega}{\omega}$ over the inter-cell coupling $K$:

$$\frac{\delta X}{X} \sim \frac{1}{K} \cdot \frac{\delta\omega}{\omega}$$

It should be noted that, during the tuning process, field flatness is achieved only for the operating mode, but for HOMs the cell resonance frequencies may differ, resulting in a significant value of $\frac{\delta\omega}{\omega}$. For very small $K$ (flat bandwidth), it is easy to obtain $\frac{\delta X}{X}$>1 and consequently, these are no longer conventional standing-wave (SW) modes in a multi-cell cavity, but rather individual cell resonances. This is exactly what was observed for the HB650 cavities, see Fig.1 and Fig.2.

Therefore, this simple model allows us to draw two important conclusions regarding a multi-cell cavity composed of pillbox cells coupled via the beam apertures on the cavity axis:

1. If the modes $TM_{0n0}$ and $TM_{0m2}$ degenerate, they form two normal modes with different frequencies and different field distribution.
2. One of the normal modes has very small electric and magnetic fields near the coupling aperture, where the partial modes mutually compensate each other. This, in turn. leads to very weak coupling between the cells, i.e., "flat" dispersion curve. Such a flat passband in a multi-cell cavity result in individual cell resonances rather than classical SW modes, and, at the same time, such a structure can exhibit a high $R/Q$ and a high loaded quality factor.
3. To first approximation, this compensation does not depend on the aperture radius $a$, and for a fixed cell length, one could simply use an enlarged beam aperture to increase the inter-cell coupling. However, this approach may prove impractical, as it leads to an increase in the electric field enhancement factor in real cavities.
4. To reduce mode interaction, mode degeneracy should be eliminated and the partial modes detuned. According to (12) the relative frequency difference has to exceed the coupling between the modes, i.e.,
$$|\omega_\alpha - \omega_\beta|/\sqrt{\omega_\alpha \omega_\beta} >> z_{\alpha\beta} z_{\beta\alpha}.$$

One way to mitigate this effect is to use a different cell length, or $\beta_G$, if permitted the beam dynamics .

## IV. MODE INTERACTION IN AN ELLIPTICAL SRF CAVITY

Let us consider an axisymmetric elliptical cavity, making the following general assumptions:

1. The cavity cells are symmetrical left-to-right.
2. There are two degenerate *axisymmetric* modes with an even variation in the longitudinal direction, namely $TM_{0n0}$ and $TM_{0m2}$.
3. The cells have equal apertures, which are small compared to the wavelengths $\lambda_{0n0} = \lambda_{0m2}$.

The elliptical cell model is shown in Fig. 8. The ideal cell with volume $V$ has no coupling apertures, and its inner surface $S$ is outlined in black. The coupling slots are treated as a perturbation. They have surface $S_1$ outlined in blue and volume $\Delta V$ (shaded area). Taking

* Contact author: yakovlev@fnal.gov

$\vec{H}_\alpha = \vec{H}_{0m2}$ and $\vec{H}_\beta = \vec{H}_{0n0}$, we can use equation (1) for the total field.

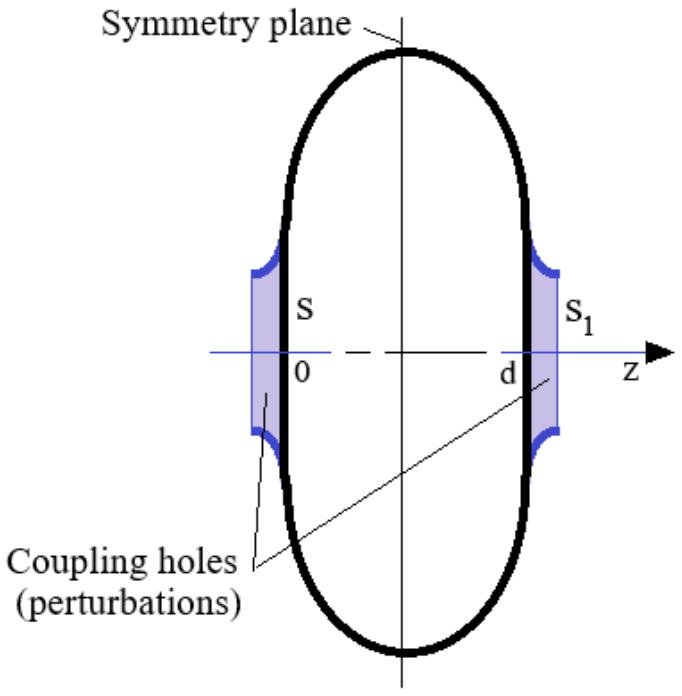

**Figure 8:** The cell geometry of an axisymmetric elliptical SRF cavity. The cell has left-to-right symmetry.

Because both partial modes are axisymmetric, near the axis the components $E_{\alpha z}$ and $E_{\beta z}$ are independent of the radius $r$ in the first approximation. At the cavity ends, i.e., for $z = 0$ and $z = d$, see Fig. 8, one has $E_{\alpha z} \approx E_{\alpha 0}$, $E_{\beta z} \approx E_{\beta 0}$ for $r \ll \lambda_{0m2} = \lambda_{0n0}$. The magnetic field there is weak and can be neglected. Due to mode degeneracy, $\omega_\alpha = \omega_\beta = \omega_0$, and, in this case, the quantities $z_{pq}$ can be estimated as follows, neglecting the magnetic field:

$$z_{\alpha\alpha} = \frac{\varepsilon_0 E_{\alpha 0}^2 \cdot \Delta V}{\mu_0 \int_V |\vec{H}_\alpha|^2 dV},$$

$$z_{\beta\beta} = \frac{\varepsilon_0 E_{\beta 0}^2 \cdot \Delta V}{\mu_0 \int_V |\vec{H}_\beta|^2 dV}, \qquad (25)$$

$$z_{\alpha\beta} = \frac{\varepsilon_0 E_{\alpha 0} E_{\beta 0} \cdot \Delta V}{\mu_0 \int_V |\vec{H}_\alpha|^2 dV},$$

$$z_{\beta\alpha} = \frac{\varepsilon_0 E_{\beta 0} E_{\alpha 0} \cdot \Delta V}{\mu_0 \int_V |\vec{H}_\beta|^2 dV},$$

Note that we again take into account that the perturbation is *convex*, see above. From (25) it follows that

$$z_{\alpha\alpha} z_{\beta\beta} = z_{\alpha\beta} z_{\beta\alpha}. \qquad (26)$$

From (12) and (26) one has the same expression for frequencies of the normal modes as for the pillbox cell, see (20):

$$\omega_\pm^2 = \omega_0^2 \left(1 + \frac{z_{\alpha\alpha} + z_{\beta\beta}}{2} \pm \frac{z_{\alpha\alpha} + z_{\beta\beta}}{2}\right) \qquad (27)$$

Here for the mode "–" the resonant frequency is equal to $\omega_0$ and, therefore, according to (13) and (25)

$$h_\beta = -h_\alpha \cdot \frac{z_{\alpha\alpha}}{z_{\alpha\beta}} = -h_\alpha \cdot \frac{E_{\alpha 0}}{E_{\beta 0}}. \qquad (28)$$

Azimuthal component of magnetic field of the eigenmodes $\alpha$ and $\beta$ next to the axis at $z = 0$ and $z = d$ can be found using the Maxwell's equation curl $\vec{H}_{\alpha,\beta} = -i\omega_0 \varepsilon_0 \vec{E}_{\alpha,\beta}$:

$$H_{\alpha,\beta,\varphi}(r) = -i\frac{r}{2}\omega_0 \varepsilon_0 E_{\alpha,\beta 0} \qquad (29)$$

And, thus, according to (28) resulting field for $z = 0$ and $z = d$ of the normal mode "–"is equal to

$$\boldsymbol{H}_\varphi(r) = -i\frac{r}{2}\omega_0 \varepsilon_0 \left(h_\alpha E_{\alpha 0} + h_\beta E_{\beta 0}\right) = 0 \; (31)$$

From the Maxwell's equation curl $\boldsymbol{H} = -i\omega_0 \varepsilon_0 \vec{\boldsymbol{E}}$ and (30) it follows that for the "–"mode at $z = 0$ and $z = d$

$$\boldsymbol{E}_z(r) = 0.$$

The radial component of the electric field of the mode "–" is also equal to zero, because the partial eigenmodes $\vec{E}_\alpha$ and $\vec{E}_\beta$ have no radial components at the ends of the ideal cell, i.e., at $z = 0$ and $z = d$.

This means that, near the coupling aperture, all electromagnetic field components of the mode "–"are equal to zero, just as was the case for pillbox cells. Consequently, all the conclusions formulated in previous section are also valid for elliptical cavities.

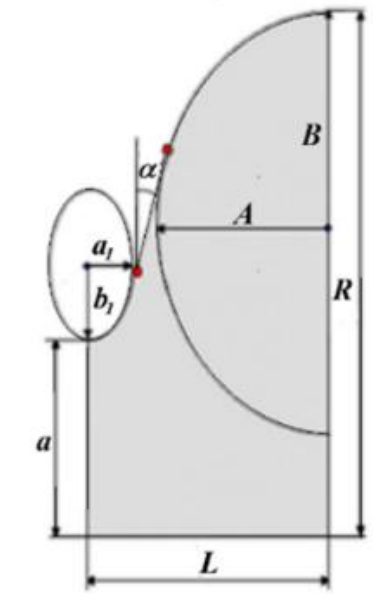

| Dimension | Regular cell | End cell |
|---|---|---|
| *a*, mm | 50 | 50 |
| *R*, mm | 200.3 | 200.3 |
| *L*, mm | 103.8 | 107.0 |
| *A*, mm | 82.5 | 82.5 |
| *B*, mm | 84 | 84.5 |
| *a₁*, mm | 18 | 20 |
| *b₁*, mm | 38 | 39.5 |
| *α*, deg | 5.2 | 7 |

(a)

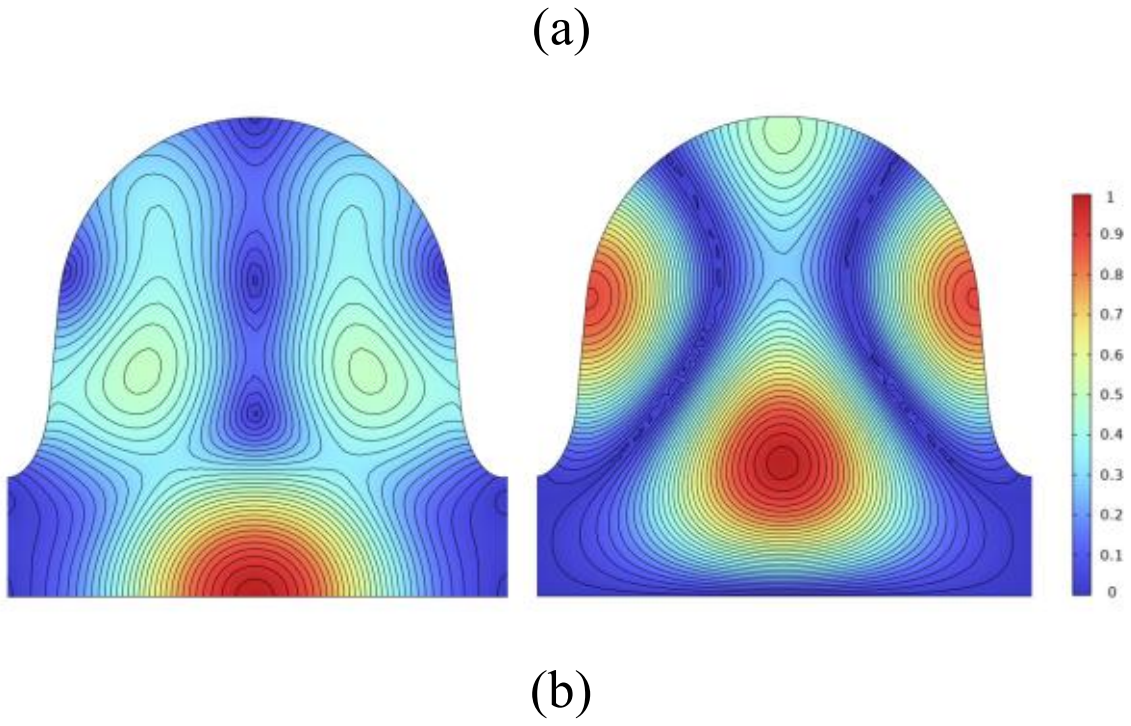

(b)

**Figure 9:** Geometric parameters of the elliptical cell for the initial PIP-II high-beta cavity design (650 MHz, $\beta_G = 0.9$) (a). Electric field distribution (left) and magnetic field distribution (right) illustrate mode mixing in the inner cell of this cavity (b).

* Contact author: yakovlev@fnal.gov

Fig. 10 shows the geometric parameters of the elliptical cell (a) and electric and magnetic field maps (b) for the hybrid modes of the 5th monopole passband of the $\beta_G = 0.9$ 650 MHz SRF cavity for the PIP-II linac[4] mentioned above. The hybrid cavity mode at 1,988 MHz results from the mixing of a pair of degenerate $TM_{030}$ and $TM_{012}$ cell modes. One can see that, in the hybrid mode, the electric and magnetic fields near to the coupling aperture are zero, and therefore, there is no coupling between the cells, which creates the trapped modes. At the same time, the cavity shape is optimized to achieve acceptable peak surface fields and low surface losses for the operating mode.

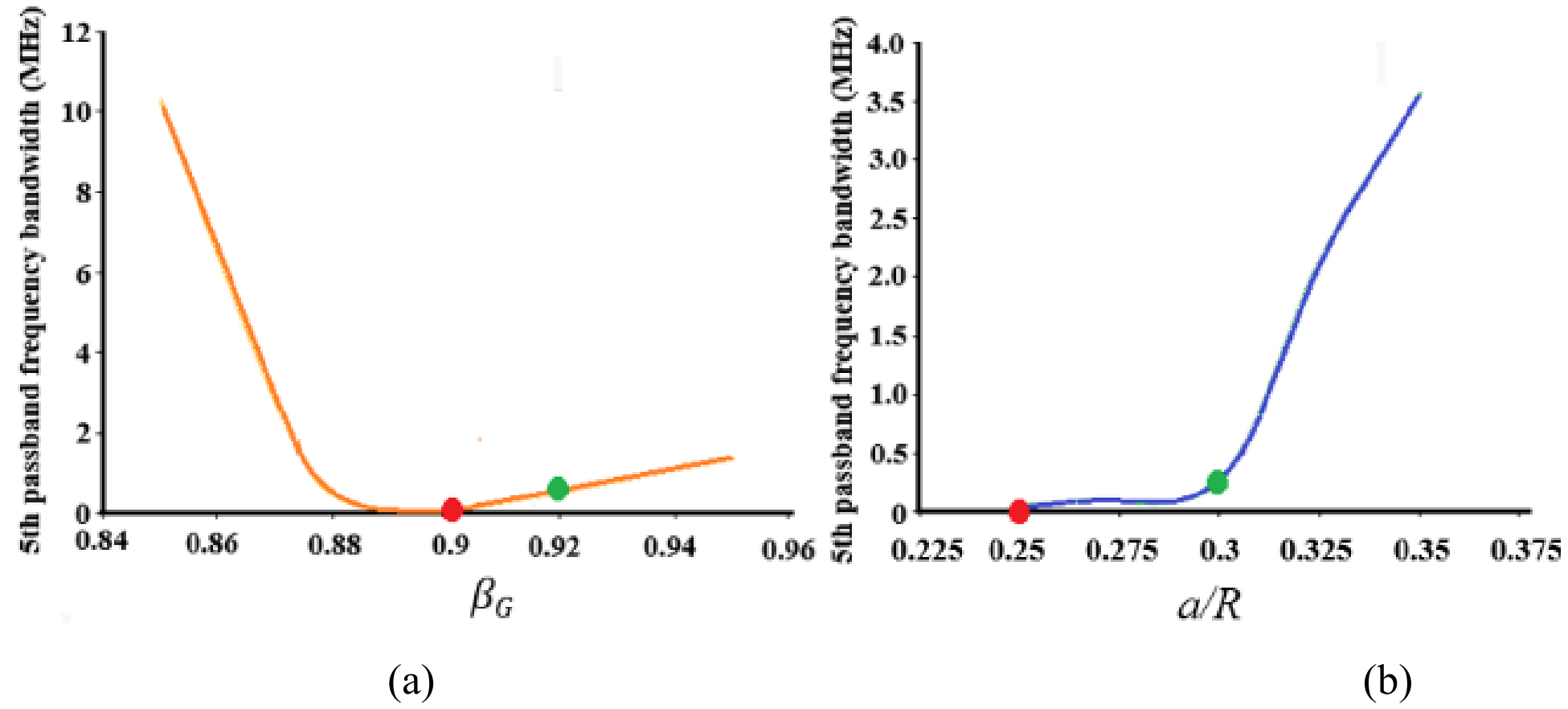

(a) (b)

**Figure 10:** Dependence of the 5th monopole passband bandwidth on the geometric beta, $\beta_G$, (a) and the normalized beam aperture (b). The red dots correspond to the initial cavity with $\beta_G = 0.9$ and an aperture radius of 50 mm. The green dots correspond to alternative cavity designs with $\beta_G = 0.92$ and an aperture radius of 50 mm (left) and with $\beta_G = 0.9$ and an aperture radius of 60 mm (right).

The frequency bandwidth $\Delta f = K f_0$ ($f_0$ is the passband average frequency) of the 5th monopole passband versus $\beta_G$ and the ratio of the aperture $a$ over the cavity equatorial radius $R$ is shown in Fig. 9 [4]. Note that the cell parameters were re-optimized for various values of $\beta_G$ and *a/R*. Nevertheless, the dependence of the bandwidth on the cavity aperture exhibits the same trend as for the pillbox case shown in Fig. 6: for the *a/R* ratio of less than 0.3 (or an aperture smaller than ~60 mm), the bandwidth - is very small, but it increases rapidly for the larger apertures. One can see that increasing $\beta_G$ to 0.92 provides a bandwidth exceeding 0.5 MHz, which is sufficient to eliminate mode splitting into individual cell resonances, as well as to achieve an acceptable $R/Q$ and a low external quality factor $Q_{ext}$. The present design of the cavity for the high-energy section of the PIP-II linac features an increased beam aperture radius of 60 mm and a parameter of $\beta_G = 0.92$. This configuration mitigates mode interaction [5] and ensures sufficient coupling for all modes of the 5th monopole passband to the input power coupler, thereby resulting in low external quality factors $Q_{ext}$ of $10^5$–$10^6$range. The cavity shape has been re-optimized to keep the peak surface fields at the levels specified in the initial design

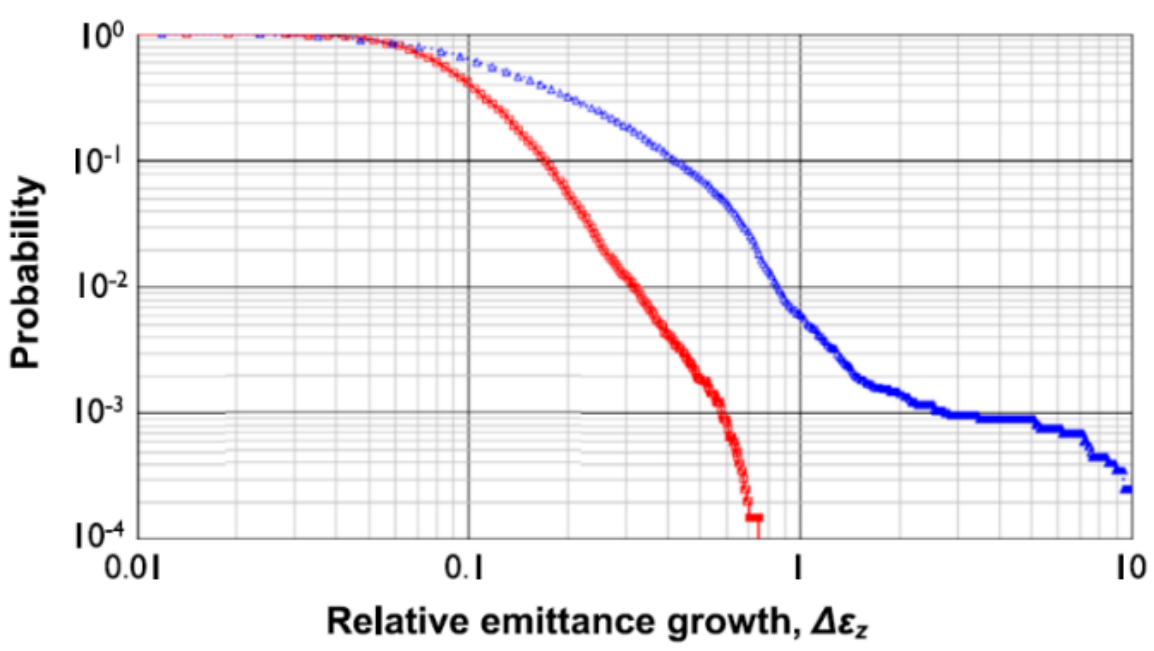

**Figure 11:** Probability of longitudinal emittance growth in the high-energy 650 MHz part of the PIP-II linac. Data for the initial design of the cavity is shown in blue, data for the present design is in red.

Using the predicted spread of the monopole HOMs parameters: resonance frequencies, $R/Q$ and $Q_{ext}$, one can generate a large number of random linac configurations (typically about $10^5$–$10^6$) to estimate the probability of longitudinal emittance growth and RF losses per cryomodule. Figure 11 shows the accumulated probability of longitudinal emittance growth in the high-energy section of the PIP-II linac at average beam current of 5 mA.

* Contact author: yakovlev@fnal.gov

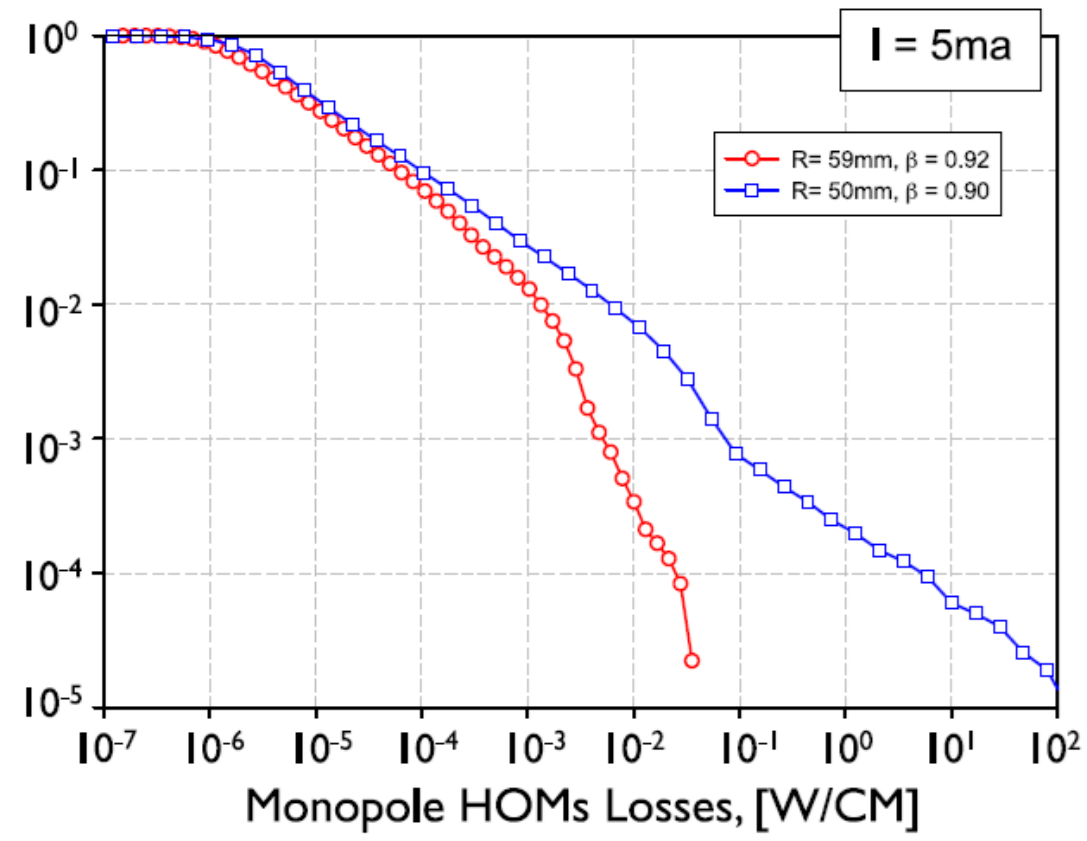

**Figure 12:** Probability of monopole HOMs RF losses per cryomodule for the initial (blue) and present (red) version of the PIP-II 650 MHz cavities. The beam current is 5 mA.

Although the probability of significant emittance increase for the initial cavity design is low,( $< 10^{-2}$), it could nevertheless limit the beam current for a future upgrade of the facility. The present design, however, does not pose problems regarding beam longitudinal emittance. The cumulative probability of HOM losses per cryomodule is presented in Fig. 12. One can see that the risk of high cryogenic losses is also significantly smaller for the present cavity design with $\beta_G = 0.92$.

## V. CONCLUSIONS

The paper describes the effect of high-order mode interaction in a multi-cell SRF accelerating structure. In the case of degenerate monopole modes, the interaction results in splitting into two normal modes. One of these modes exhibits very weak electric and magnetic fields near the beam apertures. Consequently, the cells are effectively uncoupled for this mode, yielding a nearly flat passband.

The spread of HOM frequencies due to manufacturing tolerances results in the flat passband being spit into a set of localized cell resonances that are not coupled to the beam pipe, the fundamental power coupler, or the HOM couplers. These local resonances are characterized by large shunt impedance values and may induce undesirable effects, such as beam emittance dilution and increased cryogenic losses, thus limiting the beam current. This is particularly important for the future high-current SRF linacs such as ones being developed for ADS projects [13, 14]. Therefore, this phenomenon should be carefully considered during the design and optimization of the accelerating SRF cavities.

An analytical theory based on a mode-interaction model is developed to explain this effect. Possible mitigation strategies are also discussed. It is shown that the most effective approach is to modify the cell length (i.e., the geometric beta), provided that the beam dynamics requirements permit such a change.

It is worth noting that the mode-interaction effect is deliberately exploited in the RF windows of high-power couplers to minimize both the electric and magnetic fields in the ceramic-to-metal brazing region. An example is given in Ref. [15], where the interaction between the modes $TE_{01}$ and $TE_{02}$ is described.

## ACKNOWLEDGMENTS

This work was supported by the U.S. Department of Energy, Office of Science. This manuscript has been authored by Fermi Forward Discovery Group, LLC under Contract No. DE-AC02-07CH11359 with the U.S. Department of Energy.

* Contact author: yakovlev@fnal.gov